\documentclass[9pt,twocolumn,twoside]{osajnl}
\usepackage{braket}
\usepackage{upgreek}

\journal{ol} 

\setboolean{shortarticle}{true} 

\ifthenelse{\boolean{shortarticle}}{\colorlet{color2}{color2b}}{\colorlet{color2}{color2}} 
\title{Non-reciprocal geometric phase in nonlinear frequency conversion}

\author[1,*]{Kai Wang}
\author[2]{Yu Shi}
\author[1]{Alexander S. Solntsev}
\author[2]{Shanhui Fan}
\author[1]{Andrey A. Sukhorukov}
\author[1]{Dragomir N. Neshev}

\affil[1]{Nonlinear Physics Centre, Research School of Physics and Engineering,  Australian National University, Canberra, ACT 2601, Australia}
\affil[2]{Ginzton Laboratory, Department of Electrical Engineering, Stanford University, Stanford, California 94305, USA}

\affil[*]{Corresponding author: kai.wang@anu.edu.au}

\dates{Compiled \today}

\ociscodes{(190.4410) Nonlinear optics, parametric processes; (230.3240) Isolators.}

\doi{\url{http://dx.doi.org/10.1364/OL..XX.XXXXXX}}

\begin{abstract}
We describe analytically and numerically the geometric phase arising from nonlinear frequency conversion and show that such a phase can be made non-reciprocal by momentum-dependent photonic transition. Such non-reciprocity is immune to the shortcomings imposed by dynamic reciprocity in Kerr and Kerr-like devices. We propose a simple and practical implementation, requiring only a single waveguide and one pump, while the geometric phase is controllable by the pump and promises robustness against fabrication errors.
\end{abstract}

\begin{document}

\maketitle
\thispagestyle{fancy}
\ifthenelse{\boolean{shortarticle}}{\abscontent}{}

Motivated by both fundamental aspects and practical applications, breaking the Lorentz reciprocity in optical systems is of great interest, with optical isolators~\cite{Jalas:2013-579:NPHOT} being an important example. Meanwhile, the growing interest in building optical systems on-chip has stimulated the research on new types of non-reciprocal devices, as traditional Faraday isolators require a strong magnetic field that is difficult to achieve with magneto-optical materials on-chip~\cite{Stadler:2014-600215:IPJ}. Among various effects in non-reciprocal devices, controllable non-reciprocal phase shift promises applications such as building up direction-sensitive optical phase modulators. Utilizing the phase degree of freedom of light, while preserving the frequency and power, can facilitate a range of non-reciprocal systems. Additionally, the non-reciprocal phase can enable unidirectional transmission by employing interferometric configurations. The non-reciprocal phase was typically achieved using magneto-optical effects~\cite{Shoji:2016-13001:JOPT} or effective-magnetic-field based interferometers~\cite{Tzuang:2014-701:NPHOT}.

A particular interesting type of phase is the so-called geometric phase~\cite{Berry:1984-45:PRSA}, which arises from the geometric nature of the space a state inhabits~\cite{Aharonov:1987-1593:PRL}. After a cycle of evolution, the state can acquire an additional phase factor that reflects such a geometric property, in addition to the time integral of its instantaneous Hamiltonian~(i.e. the dynamic phase). Whereas it was originally introduced in the context of quantum mechanics, the geometric phase has since been observed in a variety of optical systems~(see, e.g., \cite{Tomita:1986-937:PRL, Bliokh:2008-30404:PRL, Wang:2016-1889:OL}). An advantage of utilizing the geometric phase lies in the robustness against parameter fluctuations~\cite{Anandan:1992-307:NAT}. However, it remains an open question whether one can directly relate geometric phase with non-reciprocal applications of light.

Beyond magneto-optical materials, a variety of time-dependent and nonlinear systems have been demonstrated to possess non-reciprocal properties~(see, e.g., Refs.~\cite{Tzuang:2014-701:NPHOT,Yu:2009-91:NPHOT, Fang:2012-153901:PRL,Lira:2012-33901:PRL, Fang:2012-782:NPHOT,Peng:2014-394:NPHYS,Fan:2012-447:SCI,Miroshnichenko:2010-63302:APL,Bender:2013-234101:PRL,Huang:2011-2267:JLT, Kang:2011-549:NPHOT, Poulton:2012-21235:OE, Li:2014-3225:NCOM, Kim:2015-275:NPHYS,Konotop:2002-235208:PRB,Longhi:2016-43822:PRA,Zheng:2016-e16072:LSA,Wang:2016-1739:PTL,Hua:2016:NCOMM, Rangelov:2016:arXiv}). In a time-dependent system, a representative class of non-reciprocity in integrated optics was based on dynamic modulation of waveguides~\cite{Tzuang:2014-701:NPHOT,Yu:2009-91:NPHOT, Fang:2012-153901:PRL,Lira:2012-33901:PRL, Fang:2012-782:NPHOT}, which resulted in effectively linear isolators whose operation did not depend on the strength of the signal. On the other hand,
nonlinear effects were extensively investigated as a means to create all-optical isolators and other non-reciprocal devices. However, as was recently found, there exists a shortcoming in Kerr and Kerr-like nonlinear optical isolators~\cite{Peng:2014-394:NPHYS,Fan:2012-447:SCI,Miroshnichenko:2010-63302:APL,Bender:2013-234101:PRL} due to the so-called dynamic reciprocity~\cite{Shi:2015-388:NPHOT}. Simply put, dynamic reciprocity occurs in certain nonlinear optical isolators when they are biased with a large forward signal, under which there exists a class of small noise signals in the operation frequency range that are reciprocal~\cite{Shi:2015-388:NPHOT}. Therefore, the application of such isolators appears to be fundamentally limited. Nevertheless, within the family of nonlinear isolators, there also exist various working principles that do not suffer from the dynamic reciprocity. One example is the non-reciprocity in Stimulated Brillouin Scattering systems~\cite{Huang:2011-2267:JLT, Kang:2011-549:NPHOT, Poulton:2012-21235:OE, Li:2014-3225:NCOM, Kim:2015-275:NPHYS}, as was pointed out in Ref.~\cite{Shi:2015-388:NPHOT}. Alternatively, nonlinear frequency conversion (see, e.g. Refs.~\cite{Longhi:2016-43822:PRA,Zheng:2016-e16072:LSA,Wang:2016-1739:PTL,Hua:2016:NCOMM,Rangelov:2016:arXiv}) constitutes another important dynamic-reciprocity-free approach, where Lorentz reciprocity is broken by an optical pump wave that converts one frequency to another. Such systems are not influenced by the limitations imposed by dynamic reciprocity, provided that the pump amplitude is non-depleted and thus quasi-static, hence the interactions of the small waves become linear-like. 

In nonlinear frequency conversion with non-depleted pump, two-level system analogy has been investigated~\cite{Suchowski:2008-63821:PRA, Rangelov:2012-45804:PRA, Suchowski:2014-333:LPR, Longhi:2016-43822:PRA}.
For such processes, as noticed and utilized in quantum computation~\cite{Langford:2011-360:NAT}, the geometric phase does exist.
In this Letter, we introduce a non-reciprocal manifestation of the geometric phase in nonlinear frequency conversion. We show that by engineering the photonic band structure of the waveguide, the phase can be directly connected to the directional condition of photonic transition and hence made non-reciprocal.

\begin{figure}[tb]
	\centering
	\includegraphics[width=0.85\columnwidth]{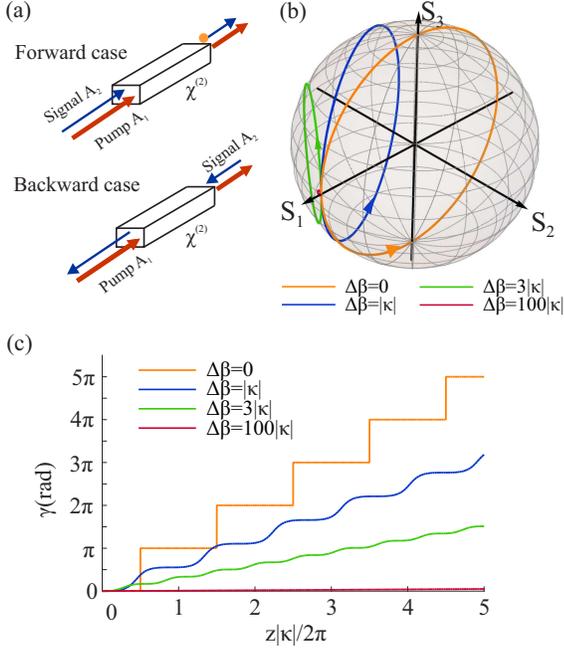}
	\caption{(a)~Sketch of a waveguide with $\chi^{(2)}$ nonlinearity for two cases of the signal direction with respect to that of the pump. (b)~The trajectory of $\ket{\tilde{\psi}}$ on the Poincar\'e sphere in four cases with different momentum mismatch $\Delta \beta=\alpha |\kappa|$. (c)~The corresponding Pancharatnam-Berry phases $\gamma$ of $\ket{\tilde{\psi}}$. }\label{fig:sketch_phase}
\end{figure}

In non-reciprocal manifestations of nonlinear frequency conversion, one typically considers two cases of the signal wave propagating in the same (forward) direction as the pump or in the opposite (backward) direction, see a sketch in Fig.~\ref{fig:sketch_phase}(a). In a waveguide geometry, the backward case is usually mismatched, such that signal propagates essentially in the linear regime. Then, we analyze in detail the effect of nonlinear interactions on the signal phase accumulation in the forward direction, considering the regime of sum-frequency-generation~(SFG) mediated by quadratic nonlinearity. We denote $\omega_1$,  $\omega_2$ and $\omega_3=\omega_1+\omega_2$ as the frequency of pump, signal and sum-frequency waves, respectively. In the slowly varying envelope approximation, we can write the electric field of each wave as
$\mathbf{E_n}\left(\mathbf{r},t\right)=A_n(z)\hat{\mathbf{E}}(x,y)\exp \left[i\left(\beta_n z-\omega_nt\right)\right]+c.c.$ for $n=1,2,3$, where $A_n(z)$ is the slowly varying envelope and $\beta_n$ is the propagation constant of each wave. If we assume that the pump is not depleted throughout the whole process, the dynamics of $a_2(z)=\sqrt{\beta_2}\omega_2^{-1}A_2(z)$ and $a_3(z)=\sqrt{\beta_3}\omega_3^{-1}A_3(z)$ can be written in terms of two differential equations~\cite{Boyd:2008:NonlinearOptics}
\begin{subequations}\label{equation:cwe2}
	\begin{align}
	i\frac{\mathrm{d}}{\mathrm{d} z}a_2(z)&=\kappa^\ast \exp\left(-i\Delta \beta z\right) a_3(z) ,\label{equation:cwe2:1}\\
	i\frac{\mathrm{d}}{\mathrm{d} z}a_3(z)&=\kappa \exp\left(i\Delta \beta z\right) a_2(z) ,\label{equation:cwe2:2}
	\end{align}
\end{subequations}
in which
$\kappa=-2d_{\mathrm{eff}}\omega_2 \omega_3 c^{-2} (\beta_2 \beta_3)^{-1/2} A_1$ with $d_{\mathrm{eff}}$ being the effective nonlinear constant, $c$ the velocity of light in vacuum, and $\Delta \beta=\beta_1+\beta_2-\beta_3$ the momentum mismatch. Due to the conservation of energy and photon number, $\left|a_2\right|^2+\left|a_3\right|^2$ is always a constant. Thus the evolution of $a_2$ and $a_3$ can be fully represented by a unitary vector $\ket{\psi}=\left[a_2,a_3\right]^{\mathrm{T}}/\sqrt{\left|a_2\right|^2+\left|a_3\right|^2}$ that spans a two dimensional Hilbert space. 
We apply a rotating-frame transformation and represent  \eqref{equation:cwe2} in the form of a Schr\"odinger-like equation:
\begin{align}\label{equation:hamilto}
i\frac{\mathrm{d}}{\mathrm{d} z}\ket{
\tilde{\psi}
}=\left[
\begin{array}{ccc}
-\Delta \beta/2 & \kappa^\ast  \\
\kappa  & \Delta \beta/2
\end{array}
\right]\ \ket{
	\tilde{\psi}
},
\end{align}
where $\ket{
	\tilde{\psi}
}=\left[\tilde{a}_2, \tilde{a}_3\right]^{\mathrm{T}}/\sqrt{\left|a_2\right|^2+\left|a_3\right|^2}$ with $\tilde{a}_2=a_2 \exp \left(-i \Delta \beta z/2\right)$ and $\tilde{a}_3=a_3 \exp \left(i \Delta \beta z/2\right)$.
Note that the cross-coupling between the two dimensions is controlled by the pump amplitude. The geometric phase is known to exist for a model system in the form of \eqref{equation:hamilto}, which is essentially the Pancharatnam-Berry phase~\cite{Pancharatnam:1956-247:PIAS}, initially defined in the context of polarization optics. Unlike the Aharonov-Anandan phase~\cite{Aharonov:1987-1593:PRL}, the Pancharatnam-Berry phase can be defined for non-cyclic evolutions (i.e. also valid if the state is not yet back to its original). The geometric phase can be obtained by removing the dynamic phase from the total phase accumulated.
For an initial state 
$\ket{\tilde{\psi}(0)}=\left[1,0\right]^{T}$, 
it can be demonstrated that the dynamic phase acquired by 
$\ket{\tilde{\psi}}$ is $\phi_d={\Delta \beta}z / 2$. Hence the geometric phase 
is found as
\begin{align}
\begin{split}
\gamma (z)&=\arg\left[ \tilde{a}^\ast_2(0) \tilde{a}_2(z)\right]-{\Delta \beta} z / {2}
=\arg\left[ a^\ast_2(0) a_2(z)\right],
\end{split}
\end{align}
which is exactly the total phase of $a_2$. Thus the phase of $a_2$ is a pure geometric phase of 
$\ket{\tilde{\psi}}$. We use an analytical solution~\cite{Boyd:2008:NonlinearOptics} of Eq.~(\ref{equation:hamilto}) to find the geometric phase as
\begin{equation}
\gamma=\arg \left\{\left[\cos (gz)+i\Delta \beta \sin(gz) /2g\right]\exp(-{\Delta \beta} z / {2})\right\},
\end{equation}
where $g$ is a constant defined by $g=\sqrt{\Delta \beta^2+4\left|\kappa\right|^2}/2$.

For a geometric illustration of the phase $\gamma$, we can visualize the state $\ket{\tilde{\psi}}$ by a Poincar\'e sphere, where the geometric phase is exactly half of the solid angle with a surface area covered by the trajectory of the state vector on the sphere~\cite{Pancharatnam:1956-247:PIAS}. If the state does not trace out a full loop, such a surface area is formed by connecting the starting and ending points of the trajectory by the shortest geodesic arc. In Fig.~\ref{fig:sketch_phase}(b) we plot the Poincar\'e sphere, in which the coordinate of the state vector is given by $S_j=\mathrm{Tr}\left[\ket{\tilde{\psi}}\bra{\tilde{\psi}}\hat{\sigma}_j\right]$, with $\hat{\sigma}_0$ being the identity matrix and $\hat{\sigma}_j$ ($j=1,2,3$) the Pauli matrices. Governed by the Hamiltonian in \eqref{equation:hamilto}, $\ket{\tilde{\psi}}$ traces out circles on the sphere. In Fig.~\ref{fig:sketch_phase}(b), we plot the trajectories of $\ket{\tilde{\psi}}$ for different values of $\Delta \beta$. We see that in all cases $\ket{\tilde{\psi}}$ experiences cyclic evolutions. More specifically, if $\Delta \beta=0$, the state walks across the north and south poles of the sphere [i.e. points $(0,0,1)$ and $(0,0,-1)$, respectively], as shown in Fig.~\ref{fig:sketch_phase}(b). Within the first cycle, before $\ket{\tilde{\psi}}$ reaches $(-1,0,0)$, the solid angle is zero as the shortest geodesic arc is along the same way back as the trajectory. At $(-1,0,0)$, the shortest geodesic arc suddenly changes to enclosing a surface area of a hemisphere, thus the solid angle becomes $2\pi\ \mathrm{sr}$. Therefore, the geometric phase experiences jumps of $\pi$ every time $\ket{\tilde{\psi}}$ passes the singularity $(-1,0,0)$ [see Fig.~\ref{fig:sketch_phase}(c)]. In a more general context, for non-zero $\Delta \beta$, as $\Delta \beta$ increases, the plane of its trajectory is tilted to a larger degree until the circle becomes a point when $\Delta \beta$ is very large. Accordingly, the geometric phase $\ket{\tilde{\psi}}$ acquired in each cycle drops with an increasing $\Delta \beta$. When $\Delta \beta$ is much larger than $|\kappa|$, the geometric phase is close to zero.

\begin{figure}[tb]
	\centering
	\includegraphics[width=0.9\columnwidth]{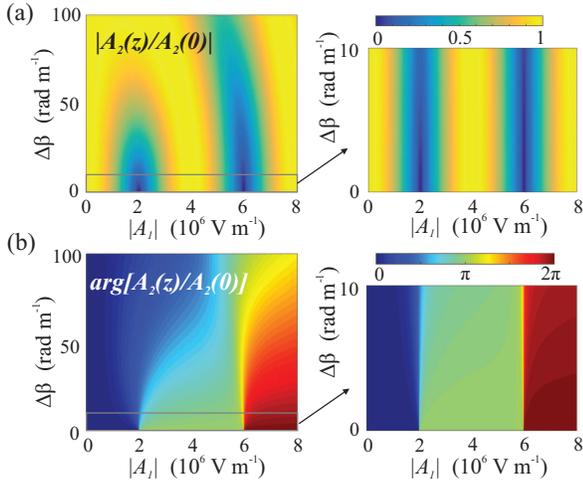}
	\caption{Estimated dependence of (a) the normalized amplitude and (b) phase of the signal $A_2(z)$ on the pump amplitude $|A_1|$ and momentum mismatch $\Delta \beta$, where different scales of $\Delta \beta$ are used for left and right plots.}
    \label{fig:mismatch}
    \vspace{-3mm}
\end{figure}

For a closer look on how the phase and amplitude depend on the momentum mismatch, we consider realistic experimental parameters: $d_{\mathrm{eff}}\sim 2.5 \times 10^{-12}\ \mathrm{m\ V^{-1}}$ (which is similar to lithium niobate crystal) and vacuum wavelengths of the pump and signal waves being $\lambda_1=1700$ nm ($\omega_1=1.11\times 10^{15}\  \mathrm{rad\ s^{-1}}$) and $\lambda_2=1200$ nm ($\omega_2=1.57\times 10^{15}\  \mathrm{rad\ s^{-1}}$), respectively. These two waves will generate a sum-frequency wave at $\omega_3=\omega_1+\omega_2=2.68\times 10^{15}\  \mathrm{rad\ s^{-1}}$ ($\lambda_3=703.45$ nm). We estimate the propagation constants based on the bulk refractive index of lithium niobate, according to the dispersion given in Ref.~\cite{Zelmon:1997-3319:JOSB}, and obtain $\beta_2 \approx 1.13\times 10^{7}\ \mathrm{rad\ m^{-1}}$ and $\beta_3 \approx 1.95\times 10^{7}\  \mathrm{rad\ m^{-1}}$. We also fix the propagation distance $z=50$ mm. Then, we calculate the dependence of the amplitude and phase of $A_2$ on $|A_1|$ and $\Delta \beta$, see Fig.~\ref{fig:mismatch}.

We observe in Fig.~\ref{fig:mismatch}(a) that when $\Delta \beta$ is small, the amplitude of the signal will have an approximately periodic dependence on the pump amplitude. The first revival of $|A_2|$ occurs at the pump amplitude of $|A_1|\approx 4.0\times 10^{6}\ \mathrm{V\ m^{-1}}$. For an effective mode area $S=1\times 10^{-12}\ \mathrm{m^{2}}$ (e.g., a square mode profile of $1\, \upmu$m size), such amplitude corresponds to a pump power of $P_1=\epsilon_0 c n_1 S |A_1|^2/2\approx 45\ \mathrm{mW}$, which is practically feasible. Importantly, $\Delta \beta$ does not have to be exactly zero for $A_2$ to exhibit revivals and $\pi$ phase shifts, and a range of small $|\Delta \beta| < 10 \ \mathrm{rad\ m^{-1}}$ could be used.

\begin{figure}[tb]
	\centering
	\includegraphics[width=.85\columnwidth]{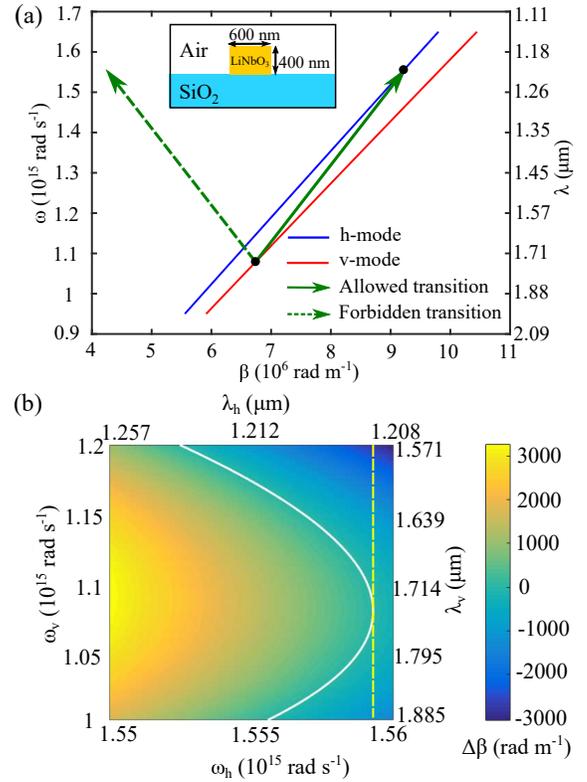}
	\caption{(a)~Schematic of a single lithium niobate nano-waveguide and the corresponding band structure for h- and v- modes, where only one-way transition is allowed.  (b)~Dependence of the momentum mismatch $\Delta \beta$ on the frequency of h-mode $\omega_h$ and that of v-mode $\omega_v$, where the white curve corresponds to the case $\Delta \beta =0$ and the dashed yellow curve denotes the case when $\omega_h=1.559\times 10^{15}\ \mathrm{rad\ s^{-1}}$.}
    \label{fig:waveguide}
    \vspace{-3mm}
\end{figure}

Now we establish how one can construct direction-dependent $\Delta \beta$ to make the geometric phase non-reciprocal. More specifically, we propose the use of photonic transitions in a nonlinear waveguide with engineered photonic band structure. We give the design of a lithium niobate nanowaveguide~\cite{Solntsev:2011-231110:APL} as an example. In Fig.~\ref{fig:waveguide}(a) we sketch the cross-section of the waveguide, with a size of $600\times 400$~nm, sitting on top of a silica substrate. We employ the two fundamental modes of horizontally and vertically polarized electro-magnetic waves (represented as h- and v- modes, respectively). Fig.~\ref{fig:waveguide}(a) plots the photonic bands (i.e. the dependency of the frequency $\omega$ on the wavevector $\beta$) for these two polarization modes. An in-band or inter-band photonic transition can happen when the interacting modes on the photonic band structure satisfy both energy and momentum conservation during nonlinear frequency conversion. In such a waveguide, the photonic band structure in Fig.~\ref{fig:waveguide}(a) permits an interband transition from a point on the band of v-mode ($\omega=1.079\times 10^{15}\ \mathrm{rad\ s^{-1}}$, $\lambda = 1747 \ \mathrm{nm}$ and $\beta=6.74 \times 10^{6}\  \mathrm{rad\ m^{-1}}$) to a point of the h-mode ($\omega=1.559\times 10^{15} \mathrm{rad\ s^{-1}}$, $\lambda = 1209 \ \mathrm{nm}$ and $\beta=9.25 \times 10^{6}\  \mathrm{rad\ m^{-1}}$), where the transition is achieved by an h-mode signal.

If the pump is prepared in v-mode to drive the transitions between the signal and the sum-frequency in h-mode, the forward and backward cases in Fig.~\ref{fig:sketch_phase}(a) can be represented by the sign of $\beta$, with $\beta>0$, for the forward case and $\beta<0$, for the other. As shown in Fig.~\ref{fig:waveguide}(a), the transition is only permitted for $\beta>0$ (green solid arrow), where $\Delta \beta \approx 0$ is expected, corresponding to the orange trajectory in Fig.~\ref{fig:sketch_phase}(b). Hence $a_2$ will pick up a $\pi$ phase shift after one cycle, as denoted by an orange dot in Fig.~\ref{fig:sketch_phase}(a). For the case $ \beta<0$, there will be $|\Delta \beta| \gg 0$, hence no photonic transition is allowed [green dashed arrow in Fig.~\ref{fig:waveguide}(a)]. This case corresponds to a fast cyclic evolution of $\ket{\tilde{\psi}}$, localized at the starting point in Fig.~\ref{fig:sketch_phase}(b), thus the solid angle is always zero and hence there will be no observable geometric phase. Note that the second-harmonic-generation (SHG) is prohibited by this photonic band structure, since for the frequency range plotted in Fig.~\ref{fig:waveguide}(a) the momentum mismatch for SHG is larger than $5\times 10^{6}\ \mathrm{rad\ m^{-1}}$ for both h- and v- modes.

In a more general context, in Fig.~\ref{fig:waveguide}(b) we also plot the dependence of momentum mismatch $\Delta \beta$ on the angular frequencies of h- and v- modes ($\omega_h$ and $\omega_v$, respectively). One can see that the momentum matching condition can be fulfilled by choosing the correct frequency of the interacting waves in the waveguide. Such a dispersion can also be utilized for achieving nonlinear frequency conversion with a certain bandwidth. For example, if one uses a pump polarized in h-mode at $\omega_1=\omega_h=1.559\times 10^{15}\ \mathrm{rad\ s^{-1}}$ [$\lambda_1=1209\ \mathrm{nm}$, dashed yellow line in Fig.~\ref{fig:waveguide}(b)], then for a signal frequency $\omega_2=\omega_v$ between $1.072\times 10^{15}\ \mathrm{rad\ s^{-1}}$ ($1758\ \mathrm{nm}$) and $1.087\times 10^{15}\ \mathrm{rad\ s^{-1}}$ ($1734\ \mathrm{nm}$) the mismatch is $\Delta \beta<10\ \mathrm{rad\ m^{-1}}$, corresponding to a bandwidth of about 24~nm.

Note that the value of the geometric phase can be controlled by the pump amplitude and frequency. For example, in Fig.~\ref{fig:waveguide}(b) one can implement different trajectories to achieve different values of the geometric phase. Such a property can be utilized in non-reciprocal phase modulators. Moreover, the geometric nature of the phase promises robustness against fabrication errors (i.e. fluctuations of $\Delta \beta$ around zero). For cyclic evolutions, the geometric phase only depends on the solid angle that the trajectory encloses on the Poincar\'e sphere. Although a fluctuating $\Delta \beta$ changes the trajectory locally, the solid angle remains stable and hence the geometric phase would stay close to $\pi$. 

On the aspect of experimental realization, the non-reciprocal geometric phase is not unique to $\chi^{(2)}$ nonlinearity. In $\chi^{(3)}$ medium, for instance, if one uses a four-wave-mixing process with a non-depleted pump and a frequency-degenerate signal, the process is SFG-like and thus two-level-system analogy in \eqref{equation:hamilto} can also be obtained, which was used to achieve quantum gates for computation~\cite{Langford:2011-360:NAT} with a single pump.
Furthermore, the non-reciprocal geometric phase is not necessarily implemented by dispersion engineered photonic system. More generally, many other systems with tailorable momentum-matching condition may be utilized. For example, in periodically poled lithium niobate waveguides, quasi-phase-matching can be used without requiring different polarizations for the interacting waves. For mode-matching systems such as all-dielectric resonance nanostructures~\cite{Kuznetsov:2016-846:SCI}, asymmetric modes in nonlinear excitation may be utilized to introduce such non-reciprocity.
In this Letter, the non-reciprocity is demonstrated in the form of phase, yet it can be easily converted to non-reciprocal transmission. As an example, a Mach-Zehnder interferometer can transform the $\pi$ non-reciprocal phase to directional transmission that bypasses the dynamic reciprocity, to construct a nonlinear isolator.

In conclusion, we demonstrate a manifestation of the geometric phase arising from the nonlinear frequency conversion. Unlike electric signal modulated systems, photonic transition is achieved and controlled by light, allowing for the realization of non-reciprocity on all-optical integrated platforms. Meanwhile, our scheme only requires a single pump and thus promises simplicity in implementation. Moreover, our results also demonstrate an important way to overcome the dynamic reciprocity in nonlinear systems. For future work, we can consider improving the bandwidth of the waveguide by engineering its dispersion. Alternatively, it may be interesting to explore the existence of non-reciprocal geometric phases in difference-frequency-generation processes.   

\paragraph{Funding.}
Australian Research Council (Discovery Project DP160100619 and Centre of Excellence CUDOS); United States Air Force Office of Scientific Research (AFOSR) Grants No. FA9550-15-1-0335, FA9550-17-1-0002.

\paragraph{Acknowledgment.}
We gratefully thank Yuri~S.~Kivshar and Benjamin~Eggleton for fruitful discussions.

\bibliography{db_Non-reciprocal_geometric_phase_in_nonlinear_frequency_conversions}

\begin{thebibliography}{10}
\newcommand{\enquote}[1]{``#1''}

\bibitem{Jalas:2013-579:NPHOT}
D.~Jalas, A.~Petrov, M.~Eich, W.~Freude, S.~H. Fan, Z.~F. Yu, R.~Baets,
  M.~Popovic, A.~Melloni, J.~D. Joannopoulos, M.~Vanwolleghem, C.~R. Doerr, and
  H.~Renner, \enquote{What is - and what is not - an optical isolator,} Nature
  Photonics \textbf{7}, 579--582 (2013).

\bibitem{Stadler:2014-600215:IPJ}
B.~J.~H. Stadler and T.~Mizumoto, \enquote{Integrated magneto-optical materials
  and isolators: A review,} IEEE Photonics J. \textbf{6}, 0600215--15 (2014).

\bibitem{Shoji:2016-13001:JOPT}
Y.~Shoji, K.~Miura, and T.~Mizumoto, \enquote{Optical nonreciprocal devices
  based on magneto-optical phase shift in silicon photonics,} J. Opt.
  \textbf{18}, 013001--14 (2016).

\bibitem{Tzuang:2014-701:NPHOT}
L.~D. Tzuang, K.~Fang, P.~Nussenzveig, S.~H. Fan, and M.~Lipson,
  \enquote{Non-reciprocal phase shift induced by an effective magnetic flux for
  light,} Nature Photonics \textbf{8}, 701--705 (2014).

\bibitem{Berry:1984-45:PRSA}
M.~V. Berry, \enquote{Quantal phase-factors accompanying adiabatic changes,}
  Proc. R. Soc. London Ser. A-Math. Phys. Eng. Sci. \textbf{392}, 45--57
  (1984).

\bibitem{Aharonov:1987-1593:PRL}
Y.~Aharonov and J.~Anandan, \enquote{Phase-change during a cyclic quantum
  evolution,} Phys. Rev. Lett. \textbf{58}, 1593--1596 (1987).

\bibitem{Tomita:1986-937:PRL}
A.~Tomita and R.~Y. Chiao, \enquote{Observation of berry topological phase by
  use of an optical fiber,} Phys. Rev. Lett. \textbf{57}, 937--940 (1986).

\bibitem{Bliokh:2008-30404:PRL}
K.~Y. Bliokh, Y.~Gorodetski, V.~Kleiner, and E.~Hasman, \enquote{Coriolis
  effect in optics: Unified geometric phase and spin-hall effect,} Phys. Rev.
  Lett. \textbf{101}, 030404--4 (2008).

\bibitem{Wang:2016-1889:OL}
K.~Wang, S.~Weimann, S.~Nolte, A.~Perez-Leija, and A.~Szameit,
  \enquote{Measuring the aharonov-anandan phase in multiport photonic systems,}
  Opt. Lett. \textbf{41}, 1889--1892 (2016).

\bibitem{Anandan:1992-307:NAT}
J.~Anandan, \enquote{The geometric phase,} Nature \textbf{360}, 307--313
  (1992).

\bibitem{Yu:2009-91:NPHOT}
Z.~F. Yu and S.~H. Fan, \enquote{Complete optical isolation created by indirect
  interband photonic transitions,} Nature Photonics \textbf{3}, 91--94 (2009).

\bibitem{Fang:2012-153901:PRL}
K.~J. Fang, Z.~F. Yu, and S.~H. Fan, \enquote{Photonic aharonov-bohm effect
  based on dynamic modulation,} Phys. Rev. Lett. \textbf{108}, 153901--5
  (2012).

\bibitem{Lira:2012-33901:PRL}
H.~Lira, Z.~F. Yu, S.~H. Fan, and M.~Lipson, \enquote{Electrically driven
  nonreciprocity induced by interband photonic transition on a silicon chip,}
  Phys. Rev. Lett. \textbf{109}, 033901--5 (2012).

\bibitem{Fang:2012-782:NPHOT}
K.~J. Fang, Z.~F. Yu, and S.~H. Fan, \enquote{Realizing effective magnetic
  field for photons by controlling the phase of dynamic modulation,} Nature
  Photonics \textbf{6}, 782--787 (2012).

\bibitem{Peng:2014-394:NPHYS}
B.~Peng, {\c{S}}.~K. {\"O}zdemir, F.~Lei, F.~Monifi, M.~Gianfreda, G.~L. Long,
  S.~Fan, F.~Nori, C.~M. Bender, and L.~Yang, \enquote{Parity-time-symmetric
  whispering-gallery microcavities,} Nature Physics \textbf{10}, 394--398
  (2014).

\bibitem{Fan:2012-447:SCI}
L.~Fan, J.~Wang, L.~T. Varghese, H.~Shen, B.~Niu, Y.~Xuan, A.~M. Weiner, and
  M.~H. Qi, \enquote{An all-silicon passive optical diode,} Science
  \textbf{335}, 447--450 (2012).

\bibitem{Miroshnichenko:2010-63302:APL}
A.~E. Miroshnichenko, E.~Brasselet, and Y.~S. Kivshar, \enquote{Reversible
  optical nonreciprocity in periodic structures with liquid crystals,} Appl.
  Phys. Lett. \textbf{96}, 063302--3 (2010).

\bibitem{Bender:2013-234101:PRL}
N.~Bender, S.~Factor, J.~D. Bodyfelt, H.~Ramezani, D.~N. Christodoulides, F.~M.
  Ellis, and T.~Kottos, \enquote{Observation of asymmetric transport in
  structures with active nonlinearities,} Phys. Rev. Lett. \textbf{110},
  234101--5 (2013).

\bibitem{Huang:2011-2267:JLT}
X.~P. Huang and S.~H. Fan, \enquote{Complete all-optical silica fiber isolator
  via stimulated brillouin scattering,} J. Lightwave Technol. \textbf{29},
  2267--2275 (2011).

\bibitem{Kang:2011-549:NPHOT}
M.~S. Kang, A.~Butsch, and P.~S. Russell, \enquote{Reconfigurable light-driven
  opto-acoustic isolators in photonic crystal fibre,} Nature Photonics
  \textbf{5}, 549--553 (2011).

\bibitem{Poulton:2012-21235:OE}
C.~G. Poulton, R.~Pant, A.~Byrnes, S.~H. Fan, M.~J. Steel, and B.~J. Eggleton,
  \enquote{Design for broadband on-chip isolator using stimulated brillouin
  scattering in dispersion-engineered chalcogenide waveguides,} Opt. Express
  \textbf{20}, 21235--21246 (2012).

\bibitem{Li:2014-3225:NCOM}
E.~B. Li, B.~J. Eggleton, K.~J. Fang, and S.~H. Fan, \enquote{Photonic
  aharonov-bohm effect in photon-phonon interactions,} Nat. Commun. \textbf{5},
  3225--5 (2014).

\bibitem{Kim:2015-275:NPHYS}
J.~Kim, M.~C. Kuzyk, K.~W. Han, H.~L. Wang, and G.~Bahl,
  \enquote{Non-reciprocal brillouin scattering induced transparency,} Nature
  Physics \textbf{11}, 275--280 (2015).

\bibitem{Konotop:2002-235208:PRB}
V.~V. Konotop and V.~Kuzmiak, \enquote{Nonreciprocal frequency doubler of
  electromagnetic waves based on a photonic crystal,} Phys Rev B \textbf{66},
  235208 (2002).

\bibitem{Longhi:2016-43822:PRA}
S.~Longhi, \enquote{Transparency in nonlinear frequency conversion,} Phys. Rev.
  A \textbf{93}, 043822--7 (2016).

\bibitem{Zheng:2016-e16072:LSA}
Y.~L. Zheng, J.~F. Yang, Z.~H. Shen, J.~J. Cao, X.~F. Chen, X.~G. Liang, and
  W.~J. Wan, \enquote{Optically induced transparency in a micro-cavity,}
  Light-Sci. Appl. \textbf{5}, e16072--5 (2016).

\bibitem{Wang:2016-1739:PTL}
K.~Wang, S.~Gao, Y.~Wang, A.~Nirmalathas, C.~Lim, K.~Alameh, and E.~Skafidas,
  \enquote{Four-wave-mixing-based silicon integrated optical isolator with
  dynamic non-reciprocity,} IEEE Photon. Technol. Lett. \textbf{28}, 1739--1742
  (2016).

\bibitem{Hua:2016:NCOMM}
S.~Hua, J.~Wen, X.~Jiang, Q.~Hua, L.~Jiang, and M.~Xiao, \enquote{Demonstration
  of a chip-based optical isolator with parametric amplification,} Nat. Commun.
  \textbf{7}, 13657--6 (2016).

\bibitem{Rangelov:2016:arXiv}
A.~Rangelov and S.~Longhi, \enquote{Nonlinear adiabatic optical isolator,}
  arXiv preprint arXiv:1612.05607  (2016).

\bibitem{Shi:2015-388:NPHOT}
Y.~Shi, Z.~F. Yu, and S.~H. Fan, \enquote{Limitations of nonlinear optical
  isolators due to dynamic reciprocity,} Nature Photonics \textbf{9}, 388--392
  (2015).

\bibitem{Suchowski:2008-63821:PRA}
H.~Suchowski, D.~Oron, A.~Arie, and Y.~Silberberg, \enquote{Geometrical
  representation of sum frequency generation and adiabatic frequency
  conversion,} Phys. Rev. A \textbf{78}, 063821--5 (2008).

\bibitem{Rangelov:2012-45804:PRA}
A.~A. Rangelov and N.~V. Vitanov, \enquote{Broadband sum-frequency generation
  using cascaded processes via chirped quasi-phase-matching,} Phys. Rev. A
  \textbf{85}, 045804--4 (2012).

\bibitem{Suchowski:2014-333:LPR}
H.~Suchowski, G.~Porat, and A.~Arie, \enquote{Adiabatic processes in frequency
  conversion,} Laser Photon. Rev. \textbf{8}, 333--367 (2014).

\bibitem{Langford:2011-360:NAT}
N.~K. Langford, S.~Ramelow, R.~Prevedel, W.~J. Munro, G.~J. Milburn, and
  A.~Zeilinger, \enquote{Efficient quantum computing using coherent photon
  conversion,} Nature \textbf{478}, 360--363 (2011).

\bibitem{Boyd:2008:NonlinearOptics}
R.~W. Boyd, \emph{{Nonlinear Optics}} (Academic Press, San Diego, 2008), 3rd
  ed.

\bibitem{Pancharatnam:1956-247:PIAS}
S.~Pancharatnam, \enquote{{Generalized theory of interference, and its
  applications.}} Proc. Indian Acad. Sci. \textbf{44}, 247--262 (1956).

\bibitem{Zelmon:1997-3319:JOSB}
D.~E. Zelmon, D.~L. Small, and D.~Jundt, \enquote{Infrared corrected sellmeier
  coefficients for congruently grown lithium niobate and 5 mol. \% magnesium
  oxide-doped lithium niobate,} J. Opt. Soc. Am. B \textbf{14}, 3319--3322
  (1997).

\bibitem{Solntsev:2011-231110:APL}
A.~S. Solntsev, A.~A. Sukhorukov, D.~N. Neshev, R.~Iliew, R.~Geiss, T.~Pertsch,
  and Y.~S. Kivshar, \enquote{Cascaded third harmonic generation in lithium
  niobate nanowaveguides,} Appl. Phys. Lett. \textbf{98}, 231110 (2011).

\bibitem{Kuznetsov:2016-846:SCI}
A.~I. Kuznetsov, A.~E. Miroshnichenko, M.~L. Brongersma, Y.~S. Kivshar, and
  B.~Luk'yanchuk, \enquote{Optically resonant dielectric nanostructures,}
  Science \textbf{354}, aag2472 (2016).

\end{thebibliography}

\clearpage
\bibliographyfullrefs{db_Non-reciprocal_geometric_phase_in_nonlinear_frequency_conversions}

\end{document}